\documentclass[journal]{IEEEtran}

\usepackage{amsmath}
\usepackage{amssymb}
\usepackage{epsfig}
\usepackage{float}
\usepackage{multirow}
\usepackage{graphicx}
\usepackage{setspace}
\usepackage{subfigure}
\usepackage{umoline}

\begin{document}

\title{Green Cellular Networks: A Survey, Some Research Issues and Challenges}

\author{Ziaul~Hasan,~\IEEEmembership{Student Member,~IEEE,}
        Hamidreza~Boostanimehr,~\IEEEmembership{Student Member,~IEEE,}\\
        and~Vijay K.~Bhargava,~\IEEEmembership{Fellow,~IEEE}}

\maketitle

\begin{abstract}
Energy efficiency in cellular networks is a growing concern for cellular operators to not only maintain profitability, but also to reduce the overall environment effects. This emerging trend of achieving energy efficiency in cellular networks is motivating the standardization authorities and network operators to continuously explore future technologies in order to bring improvements in the entire network infrastructure. In this article, we present a brief survey of methods to improve the power efficiency of cellular networks, explore some research issues and challenges and suggest some techniques to enable an energy efficient or ``green" cellular network. Since base stations consume a maximum portion of the total energy used in a cellular system, we will first provide a comprehensive survey on techniques to obtain energy savings in base stations. Next, we discuss how heterogenous network deployment based on micro, pico and femtocells can be used to achieve this goal. Since cognitive radio and cooperative relaying are undisputed future technologies in this regard, we propose a research vision to make these technologies more energy efficient. Lastly, we explore some broader perspectives in realizing a ``green" cellular network technology.
\end{abstract}

\begin{keywords}
Green communication, energy efficient networks, efficiency metrics, microcells, picocells, femtocells, cognitive radio, cooperative relaying.
\end{keywords}

\section{Introduction}
During the last decade, there has been tremendous growth in cellular networks market. The number of subscribers and the demand for cellular traffic has escalated astronomically. With the introduction of Android and iPhone devices, use of ebook readers such as iPad and Kindle and the success of social networking giants such as Facebook, the demand for cellular data traffic has also grown significantly in recent years. Hence, mobile operators find meeting these new demands in wireless cellular networks inevitable, while they have to keep their costs minimum.

\begin{figure}
\centering
\includegraphics[width=3.6in]{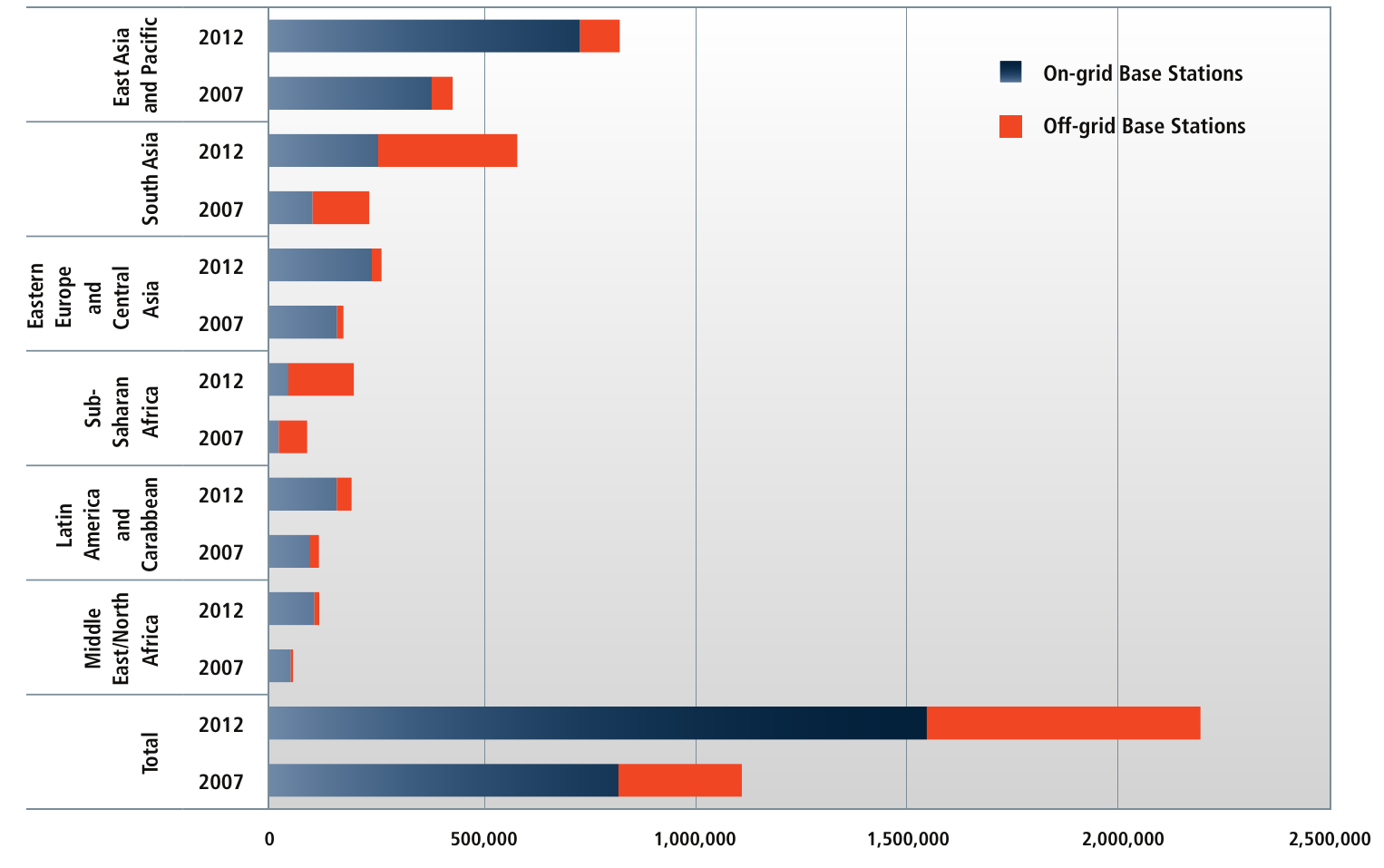}
\caption{Growth in base stations in developing regions 2007-2012 (GSMA Research) \cite{community}}
\label{growth}
\end{figure}

Such unprecedented growth in cellular industry has pushed the limits of energy consumption in wireless networks. There are currently more than 4 million base stations (BSs) serving mobile users, each consuming an average of 25MWh per year. The number of BSs in developing regions are expected to almost double by 2012 as shown in Fig. \ref{growth}. Information and Communication Technology (ICT) already represents around 2\% of total carbon emissions (of which mobile networks represent about 0.2\%), and this is expected to increase every year. In addition to the environmental aspects, energy costs also represent a significant portion of network operators' overall expenditures (OPEX). While the BSs connected to electrical grid may cost approximately 3000\$ per year to operate, the off-grid BSs in remote areas generally run on diesel power generators and may cost ten times more.

\begin{figure*}
\centering{
\subfigure[Power consumption of a typical wireless cellular network \cite{chan}(ref. therein)]{\includegraphics[width=3.4in]{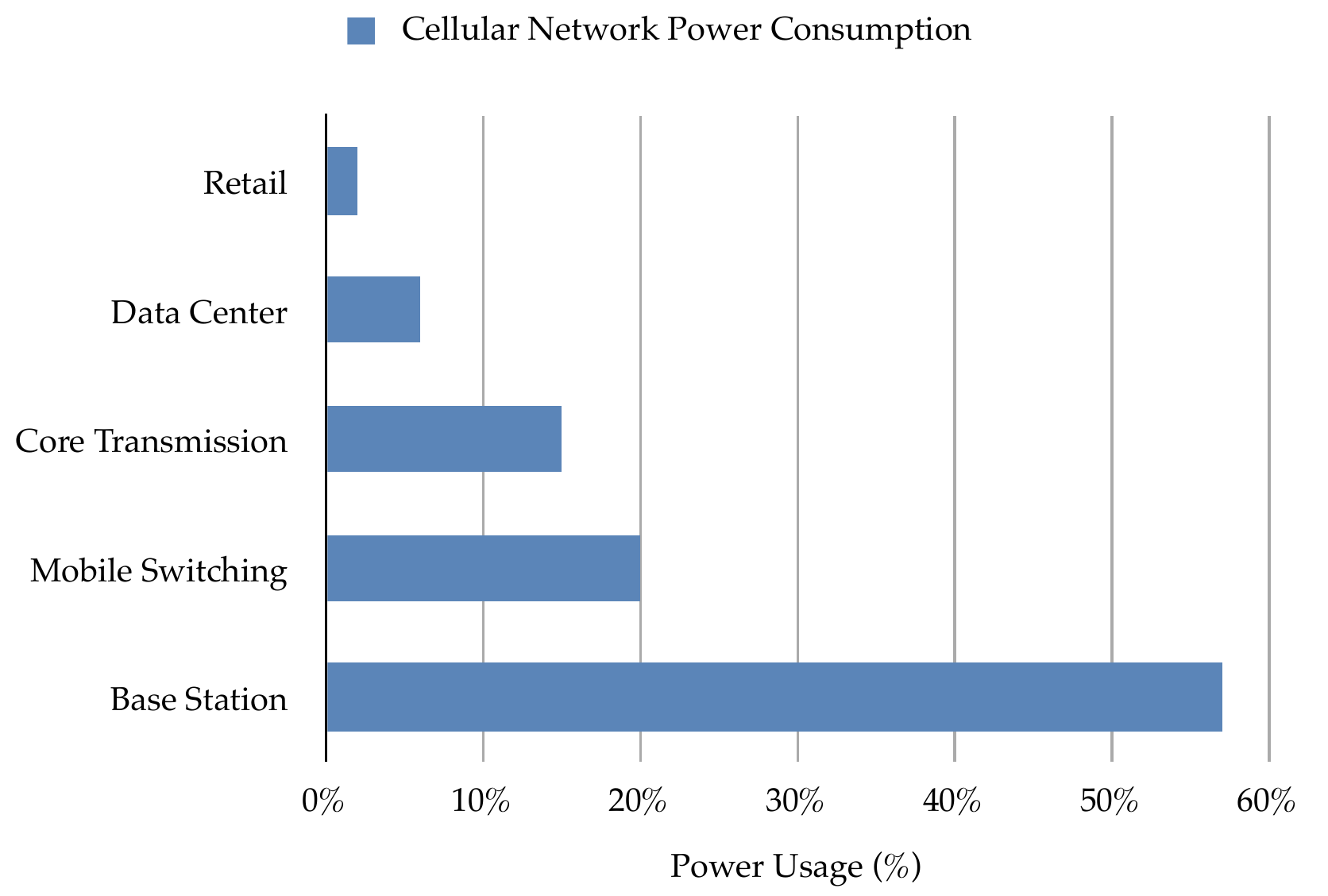}
\label{bar}}
\subfigure[Power consumption distribution in radio base stations \cite{cor}(ref. therein)]{\includegraphics[width=3.4in]{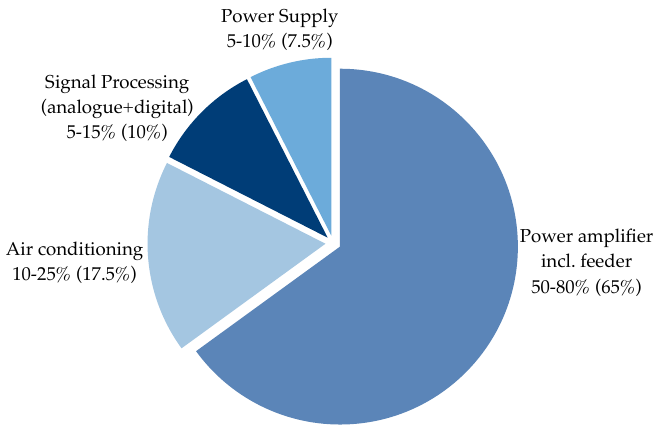}
\label{pie}}
 }
\caption{Breakdown of power consumption in a typical cellular network and corresponding base stations }
\label{barpie}
\end{figure*}

The rising energy costs and carbon footprint of operating cellular networks have led to an emerging trend of addressing energy-efficiency amongst the network operators and regulatory bodies such as 3GPP and ITU \cite{3gpp32826, itu}. This trend has stimulated the interest of researchers in an innovative new research area called ``green cellular networks". In this regard, the European Commission has recently started new projects within its seventh Framework Programme to address the energy efficiency of mobile communication systems, viz. ``Energy Aware Radio and NeTwork TecHnologies (EARTH)", ``Towards Real Energy-efficient Network Design (TREND)" and ``Cognitive Radio and Cooperative strategies for Power saving in multi-standard wireless devices (C2POWER)"  \cite{earth, trend, c2power}. ``Green radio" is a vast research discipline that needs to cover all the layers of the protocol stack and various system architectures and it is important to identify the fundamental trade-offs linked with energy efficiency and the overall performance \cite{chenyan}. Figures \ref{bar} and \ref{pie} show a breakdown of power consumption in a typical cellular network and gives us an insight into the possible research avenues for reducing energy consumption in wireless communications. In \cite{chenyan}, the authors have identified four key trade-offs of energy efficiency with network performance;  deployment efficiency (balancing deployment cost, throughput), spectrum efficiency (balancing achievable rate), bandwidth (balancing the bandwidth utilized) and delay (balancing average end-to-end service delay). To address the challenge of increasing power efficiency in future wireless networks and thereby to maintain profitability, it is crucial to consider various paradigm-shifting technologies, such as energy efficient wireless architectures and protocols, efficient BS redesign, smart grids, opportunistic network access or cognitive radio, cooperative relaying and heterogenous network deployment based on smaller cells.

\begin{figure*}
\centering
\includegraphics[width=7.0in]{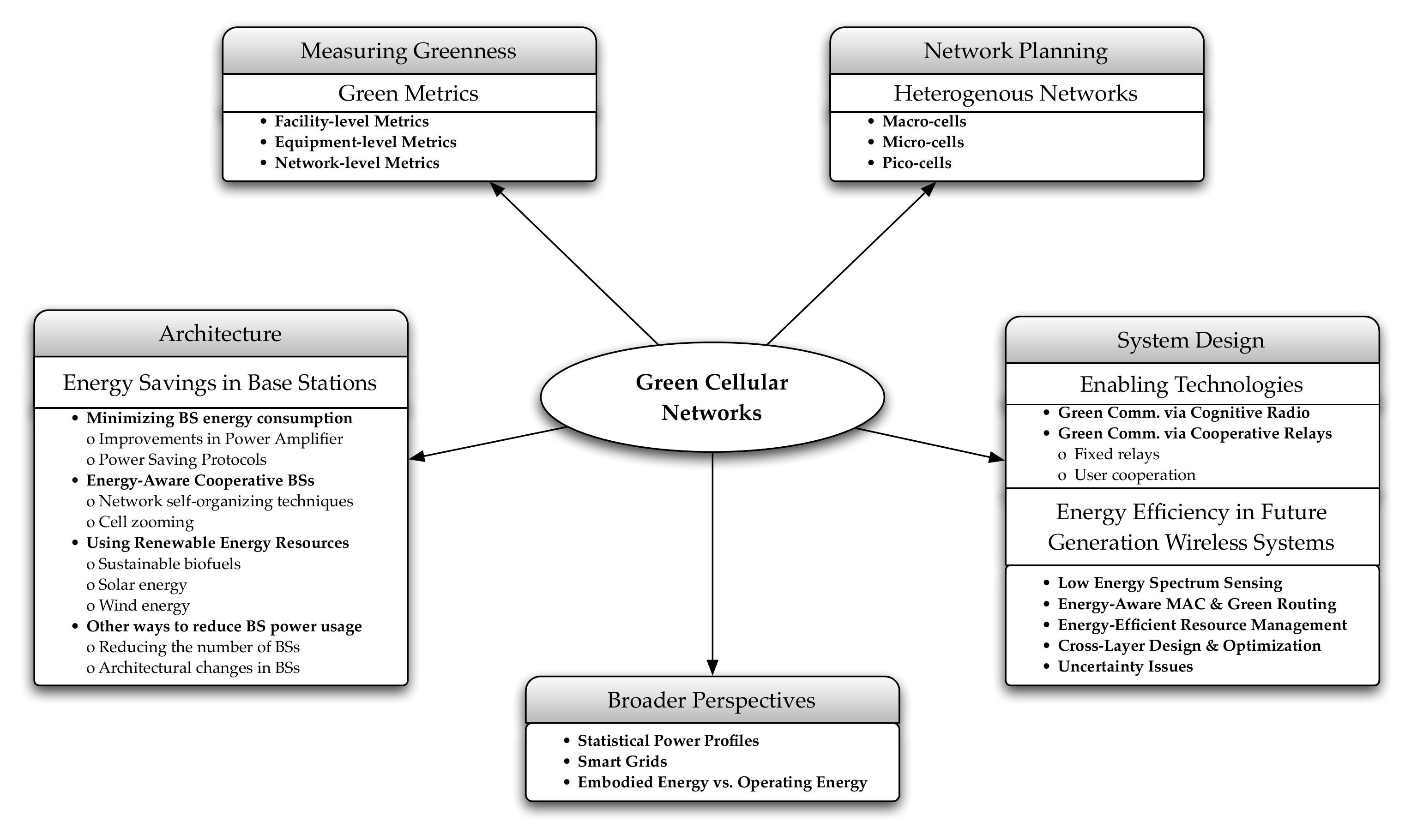}
\caption{Technical roadmap for Green Cellular Networks: A taxonomy graph}
\label{tax}
\end{figure*}

Among all the promising energy saving techniques, cognitive radio and cooperative relaying, although already getting matured in many aspects, but still are in their infancy when it comes to the deployment issues in cellular networks. Therefore, it is crucial to promote the potentials of these techniques in cellular wireless networks. Moreover, it is necessary to be aware that still many energy concerns in cognitive and cooperative networks have remained as unanswered challenges, which raises the importance of further exploring these concerns.

In this paper, we provide a brief survey on some of the work that has already been done to achieve power efficiency in cellular networks, discuss some research issues and challenges and suggest some techniques to enable an energy efficient or ``green" cellular network. We also put a special emphasis on cognitive and cooperative techniques, in order to bring attention to the benefits cellular systems can gain through employing such techniques, and also highlight the research avenues in making these techniques green. A taxonomy graph of our approach towards the design of green cellular networks is given in Fig. \ref{tax}. As shown in the figure, we identify four important aspects of a green networking  where we would like to focus: defining green metrics, bringing architectural changes in base stations, network planning, and efficient system design. In addition, some broader perspectives must also be considered. In the following sections we elaborate on each such aspect and discuss the related issues and challenges. We begin with a brief discussion on energy efficiency metrics in section \ref{metrics}. Since BSs consume the major chunk of input energy, we discuss the energy efficiency of BSs more at the component level in section \ref{eebase}. Here, we study how to minimize energy consumption of BS employing improvements in power amplifier, designing power saving protocols, implementing cooperative BS power management, using renewable energy resources and bringing some simple architectural changes. Section \ref{heter} addresses the energy efficiency from a network planning perspective where we discuss how different types of network deployments based on smaller cells can be used to increase the energy efficiency of a wireless system. Regarding the system design, we first explain the use of modern communication technologies such as cognitive radio and cooperative relays to enable green communication in cellular systems in section \ref{cogcoper} and we expand this idea further in section \ref{futureg} from a different perspective, where we discuss how the future wireless systems based on both cognitive and cooperative concepts can be made more energy efficient at the system level. Techniques such as low energy spectrum sensing, energy-aware medium access control and routing, efficient resource management, cross-layer design and addressing uncertainty issues have been examined in this context. Some broader perspectives have been discussed in \ref{broader} and conclusions are drawn in \ref{conc}.

\section{Measuring Greenness: The Metrics}
\label{metrics}
Before starting any discussion on ``green" networks, the first question naturally comes to mind is that what actually is ``green"? How do we measure and define the degree of ``greenness" in telecommunication networks? Although carbon footprint or $\textrm{CO}_2$ emissions would naturally be considered a measure of ``greenness", but the share of carbon emissions for telecommunication networks is fairly low (less than 1\%). However, please note that other motivations to obtain ``green" wireless technology also include economic benefits (lower energy costs) and better practical usage (increased battery life in mobile devices), hence evaluation of energy savings or measuring energy efficiency seems to be a more apt choice for measuring ``greenness". Thus, the notion of ``green" technology in wireless systems can be made meaningful with a comprehensive evaluation of energy savings and performance in a practical system. This is where energy efficiency metrics play an important role. These metrics provide information in order to directly compare and assess the energy consumption of various components and the overall network. In addition, they also help us to set long term research goals of reducing energy consumption. With the increase in research activities pertaining to green communications and hence in number of diverse energy efficiency metrics, standards organizations such as European Technical Standards Institute (ETSI) and Alliance for Telecommunications Industry Solutions (ATIS) are currently making efforts to define energy efficiency metrics for wireless networks \cite{atis, etsi}.

Generally speaking, energy efficiency metrics of telecommunication systems can be classified into three main categories: {\em facility-level}, {\em equipment-level} and {\em network-level} metrics \cite{aruna, tchen}. Facility-level metrics relates to high-level systems where equipment is deployed (such as datacenters, ISP networks etc.), equipment level metrics are defined to evaluate performance of an individual equipment, and network level metrics assess the performance of equipments while also considering features and properties related to capacity and coverage of the network.

The Green Grid (TGG) association of IT professionals first proposed facility-level efficiency metrics called PUE (Power Usage Efficiency) and its reciprocal DCE (Data Center Efficiency) in \cite{greengrid} to evaluate the performance of power hogging datacenters. PUE which is defined as the ratio of total facility power consumption to total equipment power consumption, is although a good metric to quickly assess the performance of datacenters at a macro level, it fails to account for energy efficiency of individual equipments. Therefore, in order to quantify efficiency at the equipment level, ratio of energy consumption to some performance measure of a communication system would be more appropriate. However, grading the performance of a communication system is more challenging than it actually first appears, because the performance comes in a variety of different forms (spectral efficiency, number of calls supported in block of time, etc.) and each such performance measure affects this efficiency metric very differently. Some suggested metrics including power per user (ratio of total facility power to number of users) measured in [Watt/user], and energy consumption rating (ECR) which is the ratio of normalized energy consumption to effective full-duplex throughput and is measured in [Watt/Gbps] \cite{ecr}.  While power per user can be a useful metric for a network provider to evaluate economic tradeoffs, network planning etc., metrics such as ECR provide the manufacturers a better insight into performance of hardware components. However,  even the busiest networks do not always operate on full load conditions, therefore it would be useful to complement metrics such as ECR to incorporate the dynamic network conditions such as energy consumption under full-load, half-load and idle cases. In this regard, other metrics such as ECRW (ECR-weighted), ECR-VL (energy efficiency metric over a variable-load cycle), ECR-EX (energy efficiency metric over extended-idle load cycle), telecommunications energy efficiency ratio (TEER) by ATIS, Telecommunication Equipment Energy Efficiency Rating (TEEER) by VerizonÕs Networks and Building Systems consider total energy consumption as weighted sum of energy consumption of the equipment at different load conditions \cite{ecr, ecr1, teer, teeer}. As an example for TEEER, the total power consumption $P_{\text{total}}$ is calculated by the following formula:
\begin{equation}
P_{\text{total}} = 0.35P_{\text{max}} + 0.4P_{50} + 0.25P_{\text{sleep}},
\label{ptotal}
\end{equation}
where $P_{\text{max}}$, $P_{50}$ and $P_{\text{sleep}}$ are power consumption at full rate, half-rate and sleep mode, respectively, and the weights are obtained statistically. However, these metrics such as ECR, TEER, TEEER etc. are unable to capture all the properties of a system and research work is still active to suggest different types of metrics. Parker {\em et al.} recently proposed an absolute energy efficiency metric (measured in dB$\varepsilon$) in \cite{parker}, given by:
\begin{equation}
\text{dB}\varepsilon=10\log_{10}\left({\text{Power/Bit Rate}\over kT\ln2}\right) ,
\end{equation}
where $k$ is the Boltzmann constant and $T$ is the absolute temperature of medium. The authors suggest that the inclusion of temperature aspect of the system is logical since classical thermodynamics is based on absolute temperature of the system under analysis. Using different examples, the authors contend that this metric is highly versatile and can be universally applied to any ICT system, subsystem and component.

While the energy efficiency metrics at the component and equipment level are fairly straightforward to define, it is more challenging to define metrics at a system or network level \cite{tchen}. Just by including the area aspect of the network, a natural choice of a metric may at first seem to be [Watt/Gbps/$\text{km}^2$], but a careful analysis can explain that it can work counter to a ``green" objective \cite{asifbhai}. Using a simple example of a typical network scenario, it has been shown in \cite{asifbhai} that due to the path loss, such a metric can only be valid when applied to networks with similar number of sites in a given area. In \cite{etsi}, ETSI proposes two network level metrics for GSM systems based on load conditions.  In rural areas, which are generally under low load conditions, the objective is to reduce power consumption in a coverage region, hence the metric is given by:
\begin{equation}
PI_{\text{rural}}={{\text{Total coverage area}}\over {\text{Power consumed at the site}}}  ,
\label{rural}
\end{equation}
where $PI_{\text{rural}}$ bears the unit of [$\text{km}^2$/Watt], and denotes the network performance indicator in rural areas. Urban areas on the hand have higher traffic demand than rural areas, hence capacity is considered instead of coverage area. A common metric under such full load conditions is therefore given by:
\begin{equation}
PI_{\text{urban}}={N_{\text{busyhour}}\over {\text{Power consumed at the site}}}  ,
\label{urban}
\end{equation}¥
where, $N_{\text{busyhour}}$ is the number of users based on average busy hour traffic demand by users and average BS busy hour traffic, and $PI_{\text{urban}}$ (users/Watt) is the network performance indicator in urban areas.

To summarize the discussion above, a non-exhaustive list of energy metrics is given in Table \ref{tb:ee}. Interested readers can find a more comprehensive taxonomy of green metrics in \cite{aruna}. Due to the intrinsic difference and relevance of various communication systems and performance measures, it is doubtful that one single metric can suffice. However, in future, the ``green" metrics must also consider deployment costs such as site construction and backhaul, and QoS requirements such as transmission delay etc. along with spectral efficiency in order to assess the true ``greenness" of the system. Once a large consensus is reached on a small set of standard energy metrics in future, it will not only accelerate the research activities in green communications, but also help pave the way towards standardization.

\begin{center}
\begin{table*}[ht]
\caption{Some energy efficiency metrics}
{\small
\hfill{}
       \begin{tabular}{ p{4cm} c c  p{7cm}}
    \hline
     \bfseries Metric     &     \bfseries Type  &     \bfseries Units    &    \bfseries Description \\ \hline
    PUE (Power Usage Efficiency) & Facility-Level & Ratio ($\ge$1) & Defined as ratio of total facility power consumption to total equipment power consumption.  \\ \hline
    DCE (Data Center Efficiency) & Facility-Level & Percentage & Defined as  reciprocal of PUE.  \\ \hline
    Telecommunications Energy Efficiency Ratio (TEER)& Equipment-Level & Gbps/Watt & Ratio of useful work to power consumption  \\ \hline
    Telecommunications Equipment Energy Efficiency Rating (TEEER) & Equipment-Level & $-\log\left({\text{Gbps}\over\text{Watt}}\right)$ & $-\log\left({P_{\text{total}}\over\text{Throughput}}\right)$, where $P_{\text{total}}$ is given by equation (\ref{ptotal}) \\ \hline
    Energy Consumption Rating (ECR) & Equipment-Level & Watt/Gbps & Ratio of energy consumption over effective system capacity  \\ \hline
    ECR-Weighted (ECRW) & Equipment-Level & Watt/Gbps & Calculated the same way as ECR except energy consumption is now calculated as $0.35E_f + 0.4E_h + 0.25E_i$, where each term corresponds to energy consumption in full load, half load and idle modes.  \\ \hline
    ECR-variable-load metric (ECR-VL) & Equipment-Level & Watt/Gbps & Average energy rating in a reference network described by an array of utilization weights \cite{ecr1}.  \\ \hline
    ECR-extended-idle metric (ECR-EX) & Equipment-Level & Watt/Gbps & Average energy rating in a reference network, where extended energy savings capabilities are enabled \cite{ecr1}.  \\ \hline
    Performance Indicator in rural areas ($PI_{\text{rural}}$) & Network-Level & $\text{km}^2$/Watt& Ratio of total coverage area to power consumed at site as given by eq. (\ref{rural})  \\ \hline
    Performance Indicator in urban areas ($PI_{\text{urban}}$) & Network-Level & users/Watt & Ratio of number of subscribers to power consumed at the site as given by eq. (\ref{urban})  \\ \hline

    \end{tabular}
}
\hfill{}
\label{tb:ee}
\end{table*}
\end{center}

\section{Architecture: Energy savings in Base Stations}
\label{eebase}
Due to the rapidly growing demand for mobile communication technology, the number of worldwide cellular BSs has increased from a few hundred thousands to many millions within a last couple of years. Such a substantial jump in the number of BSs that power a cellular network accounts for the sudden increase in greenhouse gases and pollution, in addition to higher energy costs to operate them. With the advent of data intensive cellular standards, power-consumption for each BS can increase upto 1,400 watts and energy costs per BS can reach to \$3,200 per annum with a carbon footprint of 11 tons of  $\textrm{CO}_2$ \cite{deloitte}. The radio network itself adds up to 80\% of an operator's entire energy consumption. Therefore, BS equipment manufacturers have begun to offer a number of eco and cost friendly solutions to reduce power demands of BSs and to support off-grid BSs with renewable energy resources. Nokia Siemens Networks Flexi Multiradio Base Station,  Huawei Green Base Station and Flexenclosure E-site solutions are examples of such recent efforts \cite{nokia, huawei, flexenclosure}. In \cite{louhi}, the authors present various methods dealing with improved transmitter efficiency, system features, fresh air-cooling, renewable energy sources and energy saving during low traffic. A typical cellular network consists of three main elements; a core network that takes care of switching, BSs providing radio frequency interface, and the mobile terminals in order to make voice or data connections. As the number of BSs increases, it becomes crucial to address their energy consumption for a cellular network. In the next few subsections, we will discuss different ways to reduce energy consumption due to BSs.

\subsection{Minimizing BS energy consumption}
The energy consumption of a typical BS can be reduced by improving the BS hardware design and by including additional software and system features to balance between energy consumption and performance. In order to improve hardware design of a BS for energy consumption, we need to address the energy efficiency of the power amplifier (PA). A PA dominates the energy consumption of a BS and its energy efficiency depends on the frequency band, modulation and operating environment \cite{louhi}. Some typical system features to improve BS energy efficiency are to shut down BS during low traffic or cell zooming \cite{niu, marsan}. Besides hardware redesign and new system level features, there are various site level solutions that can be used in order to save energy. For example, outdoor sites can be used over wider level of temperatures, and thus less cooling would be required. Another solution is to use more fresh air-cooling rather than power consuming air conditioners for indoor sites. In addition, RF heads and modular BS design can be implemented to reduce power loss in feeder cables \cite{louhi}.

\subsubsection{Improvements in Power Amplifier}
There are three essential parts of a BS: radio, baseband and feeder. Out of these three, radio consumes more than 80\% of a BS's energy requirement, of which power amplifier (PA) consumes almost 50\% \cite{ashwin}. Shockingly, 80-90\% of that is wasted as heat in the PA, and which in turn requires air-conditioners, adding even more to the energy costs. The total efficiency of a currently deployed amplifier, which is the ratio of AC power input to generated RF output power, is generally in anywhere in the range from 5\% to 20\% (depending on the standard viz. GSM, UMTS, CDMA and the equipment's condition) \cite{claussen}. Modern BSs are terribly inefficient because of their need for PA linearity and high peak-to-average power ratios (PAPR). The modulation schemes that are used in communication standards such as WCDMA/HSPA and LTE are characterized by strongly varying signal envelopes with PAPR that exceeds 10dB. To obtain high linearity of the PAs in order to maintain the quality of radio signals, PAs have to operate well below saturation, resulting in poor power efficiency \cite{cor}. Depending on their technology (e.g Class-AB with digital pre-distortion) and implementation, the component level efficiency of modern amplifiers for CDMA and UMTS systems is in the order of approximately 30\% to 40\% \cite{claussen}. Since these technologies have reached their limits,  PAs based on special architectures such as digital pre-distorted Doherty-architectures and GaN (Aluminum Gallium nitride) based amplifiers seem to be more promising by pushing the power efficiency levels to over 50\% \cite{claussen}. Doherty PAs that consist of a carrier and a peak amplifier is advantageous by providing easy additional linearization using conventional methods such as feed-forward and envelope elimination and restoration (EER)\cite{james}. Since GaN structures can work under higher temperature and higher voltage, they can potentially provide a higher power output. Additional improvements in efficiency can be obtained by shifting to switch-mode PAs from the traditional analog RF-amplifiers. Compared to standard analog PAs, switch-mode PAs tend to run cooler and draw less current. While amplifying a signal, a switch-mode amplifier turns its output transistors on and off at an ultrasonic rate. The switching transistors produce no current when they are switched off and produce no voltage when switched on, therefore generate very little power as heat resulting in a highly efficient power supply. It is expected that overall component-efficiency of these energy efficient devices could be around 70\% \cite{claussen}.

One more significant setback in increasing power efficiency with PAs is that they perform better at maximum output power in order to maintain the required signal quality. However, during the low traffic load conditions (e.g night time), lot of energy is routinely wasted. Therefore, design of flexible PA architectures that would allow a better adaptation of the amplifier to the required output power needs to be addressed \cite{claussen}. In addition to this, we need to investigate more efficient modulation schemes, because modulation also affects the PA efficiency. As an example, by focusing more on higher modulation schemes that require additional filtering in order to prioritize data over voice, linearity of PA is more desirable because of the non-constant envelope of the signal \cite{ashwin}. Using different linearization techniques such as Cartesian feedback, digital pre-distortion and feed-forward along with different kind of DSP methods that reduces the requirement on the linear area of PA have also been suggested \cite{louhi}.

\subsubsection{Power Saving Protocols}
In the current cellular network architecture based on WCDMA/HSPA, BSs and mobile terminals are required to continuously transmit pilot signals. Newer standards such as LTE, LTE-Advanced and WiMAX have evolved to cater ever-growing high speed data traffic requirements. With such high data requirements, although BSs and mobile units (MU) employing newer hardware (such as multiple-input and multiple-output (MIMO) antennas) increase spectral efficiency allowing to transmit more data with the same power, power consumption is still a significant issue for future high speed data networks and they require energy conservation both in the hardware circuitry and protocols. A fairly intuitive way to save power is to switch off the transceivers whenever there is no need to transmit or receive. The LTE standard utilizes this concept by introducing power saving protocols such as discontinuous reception (DRX) and discontinuous transmission (DTX) modes for the mobile handset. DRX and DTX  are methods to momentarily power down the devices to save power while remaining connected to the network with reduced throughput. Continuous transmission and reception in WCDMA/HSPA consumes significant amount of power even if the transmit powers are far below the maximum levels, and therefore power savings due to DRX and DTX is an attractive addition. IEEE 802.16e or Mobile WiMAX also has similar provisions for sleep mode mechanisms for mobile stations \cite{ltechang}. The device negotiates with the BS and the BS will not schedule the user for transmission or reception when the radio is off. There are three power-saving classes with different on/off cycles for the WiMAX standard.

Unfortunately, such power saving protocols for BSs have not been considered in the current wireless standards. The traffic per hour in a cell varies considerably over the time and BSs can regularly be under low load conditions, especially during the nighttime. In future wireless standards, energy saving potential of BSs needs to be exploited by designing protocols to enable sleep modes in BSs. The authors in \cite{cor} suggest making use of downlink DTX schemes for BSs by enabling micro-sleep modes (in the order of milliseconds) and deep-sleep modes (extended periods of time). Switching off inactive hardware of BSs during these sleep modes can potentially save a lot of power, especially under low load conditions.

\subsection{Energy-Aware Cooperative BS Power Management}
Traffic load in cellular networks have significant fluctuations in space and time due to a number of factors such as user mobility and behaviour. During daytime, traffic load is generally higher in office areas compared to residential areas, while it is the other way around during the night. Therefore, there will always be some cells under low load, while some others may be under heavy traffic load. Hence, a static cell size deployment is not optimal with fluctuating traffic conditions. For next generation cellular networks based on microcells and picocells and femtocells, such fluctuations can be very serious. While limited cell size adjustment called ``cell-breathing" currently happens in currently deployed CDMA networks (a cell under heavy load or interference reduces its size through power control and the mobile user is handed off to the neighbouring cells), a more network-level power management is required where multiple BSs coordinate together. Since operating a BS consumes a considerable amount of energy, selectively letting BSs go to sleep based on their traffic load can lead to significant amount of energy savings. When some cells are switched off or in sleep mode, the radio coverage can be guaranteed by the remaining active cells by filling in the gaps created. Such concepts of self-organizing networks (SON) have been introduced in 3GPP standard (3GPP TS 32.521) to add network management and intelligence features so that the network is able to optimize, reconfigure and heal itself in order to reduce costs and improve network performance and flexibility {\cite{3gamericas}}. The concept of SONs can be applied in order to achieve diverse objectives. For instance, in {\cite{schmelz}} different use cases for SONs are discussed, e.g., load balancing, cell outage management, management of relays and repeaters, etc. In the context of power efficiency, the performance of these self-organizing techniques were initially explored in {\cite{marsan,marsanmeo}}. Using the numerical results, the authors here suggested that substantial amount of energy savings can be obtained (of the order of 20\%, and above) by selectively reducing the number of active cells that are under low load conditions. On the other hand, a distributed algorithm is proposed in {\cite{viering}} in which BSs exchange information about their current level of power and take turns in reducing their powers. Recently, authors of {\cite{samadanis1,samadanis2}} introduced the notion of energy partitions which is the associations among powered-on and powered-off BSs, and use this notion as the basis of rearranging the energy configuration.

A similar but even more flexible concept called ``Cell Zooming" was presented in \cite{niu}. Cell zooming is a technique through which BSs can adjust the cell size according to network or traffic situation, in order to balance the traffic load, while reducing the energy consumption. When a cell gets congested with increased number of users, it can zoom itself in, whereas the neighboring cells with less amount of traffic can zoom out to cover those users that cannot be served by the congested cell. Cells that are unable to zoom in may even go to sleep to reduce energy consumption, while the neighboring cells can zoom out and help serve the mobile users cooperatively. Another such proposal to dynamically adjust cell-size in a multi-layer cellular architecture was presented in \cite{bhaumik}.

\subsubsection{Implementation}
The framework for cell zooming can include a cell-zooming server (CS) (implemented in the gateway or distributed in the BSs) that senses the network state information such as traffic, channel quality etc \cite{niu} and hence makes decisions for for cell zooming. If there is a need for a cell to zoom in or out, it will coordinate with its neighboring cells by the assistance of a CS. Cells can zoom in or out by a variety of techniques such as physical adjustment, BS cooperation and relaying \cite{niu}. Physical adjustment can be either done by adjusting the transmit powers of BSs and also by adjusting antenna height and tilt for cells to zoom in or out. BS cooperation here means that multiple BSs cooperatively transmit or receive from MUs. For an MU, a cluster of BSs cooperating form a new cell, the size of which is sum of cell sizes of these BSs. Relaying can also be used for cell size adjustment in a way that relay stations can help transfer the traffic from a cell with heavy load to a cell in low load conditions \cite{niu, bhaumik}. The authors in \cite{bhaumik} propose dynamic self-organization of the cellular layers using techniques such as timed sleep mode, user location prediction and reverse channel sensing. In such networks, BS can also go to sleep mode where the energy consuming equipments such as air conditioner etc., can be switched off. The neighboring cells can then reconfigure to guarantee the coverage.

\subsubsection{Benefits and Challenges}
Self-organizing cellular networks can be useful in load balancing as well as energy conservation by deciding when to disperse load for load balancing and when to concentrate load for energy savings. The advantages of techniques such as cell zooming also include improved user experience such as better throughput and increased battery life. For e.g in \cite{bhaumik}, a two-layer cellular architecture achieves a power savings of up to 40\% over the entire day. With techniques such as BS cooperation and relaying, inter-cell interference and fading effects can be mitigated and hence MUs can observe higher diversity gains and better coverage. However, sufficient challenges lay ahead to practically realize these networks such as radio frequency planning, configuring switching thresholds, avoiding coverage holes, tracing spatial and temporal traffic load fluctuations etc. \cite{niu, bhaumik}.

\subsection{Using Renewable Energy Resources}
In several remote locations of the world such as Africa and Northern Canada, electrical grids are not available or are unreliable. Cellular network operators in these off-grid sites constantly rely on diesel powered generators to run BSs which is not only expensive, but also generates $\textrm{CO}_2$ emissions. One such generator consumes an average of 1500 litres of diesel per month, resulting in a cost of approximately \$30,000 per year to the network operator. Moreover, this fuel has to be physically brought to the site and sometimes it is even transported by helicopter in remote places, which adds further to this cost. In such places, renewable energy resources such as sustainable biofuels, solar and wind energy seem to be more viable options to reduce the overall network expenditure. Hence, adopting renewable energy resources could save cellular companies such recurrent costs, since they are capital intensive and cheaper to maintain. Also, since renewable energy is derived from resources that are regenerative, renewable energy resources do not generate greenhouse gases such as $\textrm{CO}_2$.

Recently, a program called ``Green Power for Mobile" to use renewable energy resources for BSs has been started by 25 leading telecoms including MTN Uganda and Zain, united under the Global Systems for Mobile communications Association (GSMA) \cite{renew}. This program is meant to aid the mobile industry to deploy solar, wind, or sustainable biofuels technologies to power 118,000 new and existing off-grid BSs in developing countries by 2012. Powering that many BSs on renewable energy would save up to 2.5 billion litres of diesel per annum (0.35\% of global diesel consumption of 700 billion litres per annum) and cut annual carbon emissions by up to 6.8 million tonnes.

Such BSs operating on renewable energy resources are expensive and network operators have been reluctant to adopt them because of fear of little commercial viability and lack of equipment expertise. However, according to a bi-annual recent report by GSMA, the implementation of green power technology represents a technically feasible and financially attractive solution with a payback period of less than three years at many sites \cite{biannual}.

\subsection{Other ways to reduce BS power consumption}
Since the energy consumption of the entire cellular network includes the summation of energy used by each BS, reducing the number of BSs has a direct impact on energy consumption of a cellular network. However, efficient network design and finding an optimal balance between cell size and BS capacity can be very challenging. Features such as 2-way and 4-way diversity, feeder less site, extended cell, low frequency band, 6-sector site and smart antenna can be used to minimize the number of BS sites \cite{louhi}.

Another way to improve power efficiency of a BS is to bring some architectural changes to the BS. Currently, the connection between the RF-transmitter and antenna is done by long coaxial cables that add almost 3dB to the losses in power transmission and therefore, low power RF-cables should be used and RF-amplifier has to be kept closer to the antenna \cite{claussen}. This will improve the efficiency and reliability of the BS. In \cite{vandana}, the authors suggest an all-digital transmitter architecture for green BS that uses a combination of EER and pulse width modulation (PWM)/pulse position modulation (PPM) modulation.

\section{Network Planning: Heterogeneous Network Deployment}
\label{heter}
\begin{figure*}
\centering
\includegraphics[width=6.5in]{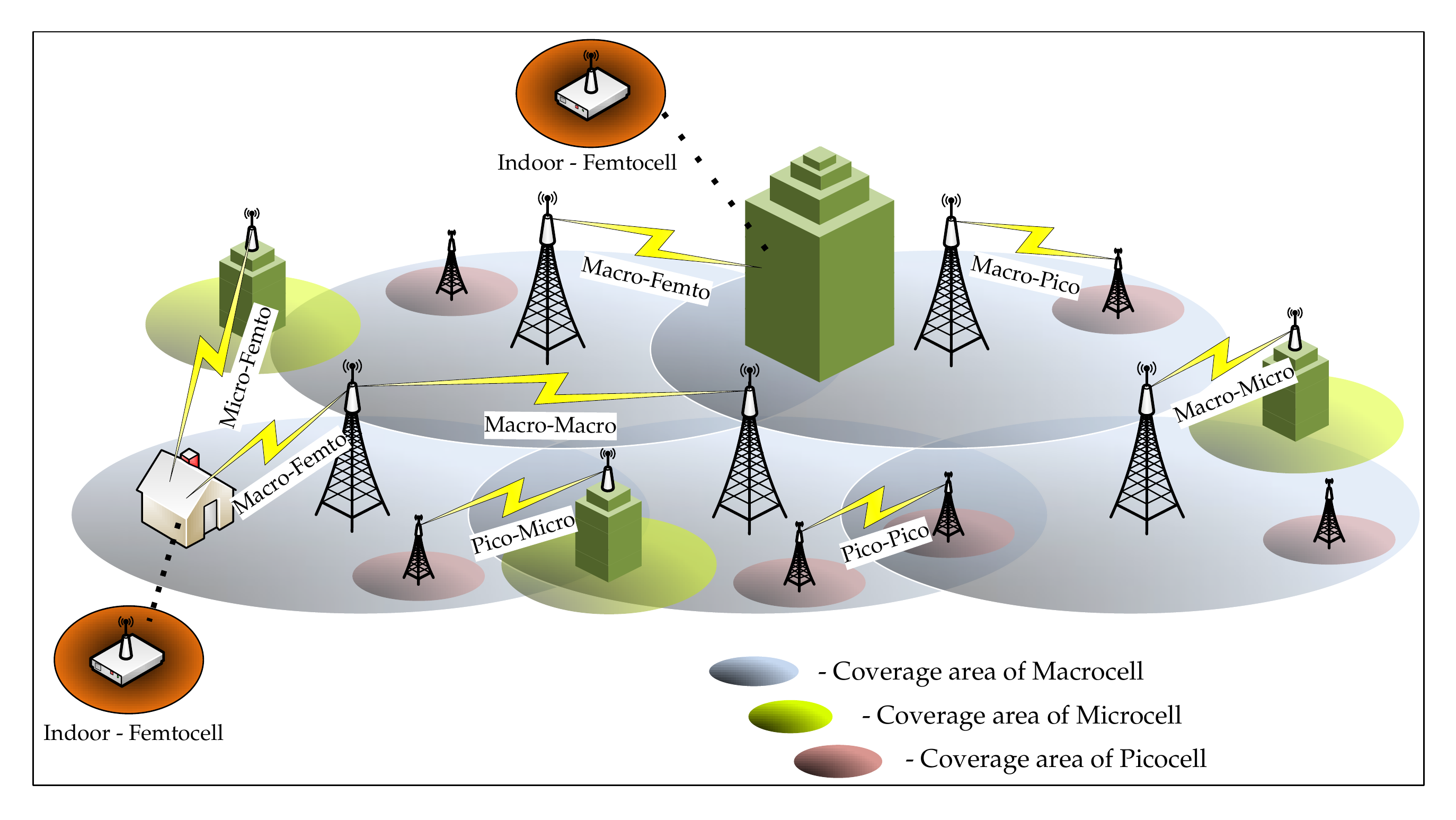}
\caption{A typical heterogeneous network deployment}
\label{hetero}
\end{figure*}

The exponential growth in demand for higher data rates and other services in wireless networks requires a more dense deployment of base stations within network cells. Whereas conventional macro-cellular network deployments are less efficient, it may not be economically feasible to modify the current network architectures. Macrocells are generally designed to provide large coverage and are not efficient in providing high data rates. One obvious way to make the cellular networks more power efficient in order to sustain high speed data-traffic is by decreasing the propagation distance between nodes, hence reducing the transmission power. Therefore, cellular network deployment solutions based on smaller cells such as micro, pico and femtocells are very promising in this context. A typical heterogeneous network deployment is shown in Fig. \ref{hetero}. A micro/picocell is a cell in a mobile phone network served by a low power cellular BS that covers a small area with dense traffic such as a shopping mall, residential areas, a hotel, or a train station. While a typical range of a micro/picocell is in the order of few hundred metres, femtocells are designed to serve much smaller areas such as private homes or indoor areas. The range of femtocells is typically only a few metres and they are generally wired to a private owners' cable broadband connection or a home digital subscriber line (DSL). Smaller cells because of their size are much more power efficient in providing broadband coverage. As an example, a typical femtocell might only have a 100mW PA, and draw 5W total compared to a 5KW that would be needed to support macrocell. An analysis by OFCOM (UK regulator) and Plextek concluded that femtocell deployment could have a 7:1 operational energy advantage ratio over the expansion of the macrocell network to provide approximately similar indoor coverage \cite{plextek}. Simulations show that with only 20\% of customers with picocells, a joint deployment of macrocell and picocell in a network can reduce the energy consumption of the network by up to 60\% compared to a network with macro-cells only \cite{claussen}. Another advantage of smaller cells is that they can use higher frequency bands suitable to provide high data rates and also offer localization of radio transmissions. However, deploying too many smaller cells within a macrocell may reduce the overall efficiency of the macrocell BS, since it will have to operate under low load conditions. Therefore, careful investigation of various deployment strategies should be done in order to find how to best deploy such smaller cells. In \cite{calin}, Calin {\em et al.} provided insight into possible architectures/scenarios for joint deployments of macro and femtocells with an analysis framework for quantifying potential macro-offloading benefits in realistic network scenarios. Richter {\em et al.} in \cite{richter1}, investigate the impact of different deployment strategies on the power consumption of mobile communication network. Considering layouts with different number of micro BSs in a cell, in addition to macro sites, the authors introduce the concept of area power consumption as a system performance metric. Simulation results suggest that under full traffic load scenarios, the use of micro BSs has a rather moderate effect on the area power consumption of a cellular network and strongly depends on the offset power consumption of both the macro and micro sites \cite{richter1}. In \cite{richter2}, the authors investigate the potential improvements of the same metric achievable in network layouts with different numbers of micro BSs together with macro sites for a given system performance targets under full load conditions.

As large-scale femtocell deployment can result in significant energy consumption, an energy saving procedure that allows femtocell BS to completely turn off its transmissions and processing when not involved in an active call was proposed in \cite{ashrafbhai}. Depending on the voice traffic model, this mechanism can provide an average power saving of 37.5\% and for a high traffic scenario, it can achieve five times reduction in the occurrence of mobility events, compared to a fixed pilot transmission \cite{ashrafbhai}. A rather radial approach to create a link between fully centralized (cellular) and decentralized (ad hoc) networks in order to achieve more efficient network deployment is a paradigm shift towards self-organizing small-cell networks (SCNs) \cite{hoydis}. However, coverage and performance prediction, interference and mobility management together with security issues are some of the many issues that must be dealt while designing such networks.

\section{Enabling Technologies: Cognitive Radio and Cooperative Relaying}
\label{cogcoper}
Recently, the research on technologies such as cognitive radio and cooperative relaying has received a significant attention by both industry and academia. While cognitive radio is an intelligent and adaptive wireless communication system that enables us to utilize the radio spectrum in a more efficient manner, cooperative relays can provide a lot of improvement in throughput and coverage for futuristic wireless networks. However, developments in both these technologies also enable us to solve the problem of energy efficiency via smart radio transmission and distributed signal processing. In the following subsections, we will discuss how we can enable green communication in cellular systems using cognitive radio and cooperative relaying.

\subsection{Green Communication via Cognitive Radio}
\label{cognitiveradio}
Bandwidth efficiency has been always a crucial concern for wireless communication engineers, and there exist a rich literature on this matter, resulting in bandwidth efficient systems, but not always considering power efficiency. On the other hand, it has been realized that the allocated spectrum is highly underutilized \cite{fcc}, and this is where cognitive radio comes into the picture. The main purpose of cognitive radio is to collect information on the spectrum usage and to try to access the unused frequency bands intelligently, in order to compensate for this spectrum underutilization \cite{haykins}. However, the question is why using spectrum more efficiently is important and how it can reduce power consumption? The answer lies under Shannon's capacity formula \cite{shannon}, where we can see the trade-off between the bandwidth and power. The capacity increases linearly with bandwidth, but only logarithmically with power. This means that in order to reduce power, we should seek for more bandwidth \cite{grace}, or in other words, manage the spectrum optimally and dynamically, and this falls into the scope of cognitive radio. In fact, it has been shown in {\cite{holland}} that up to {$50\%$} of power can be saved if the operator dynamically manages its spectrum by activities such as dynamically moving users into particularly active bands from other bands, or the sharing of spectrum to allow channel bandwidths to be increased.

However, efficient spectrum usage is not the only concern of cognitive radio. Actually, in the original definition of cognitive radio by J. Mitola \cite{mitola}, every possible parameter measurable by a wireless node or network is taken into account (Cognition) so that the network intelligently modifies its functionality (Reconfigurability) to meet a certain objective. One of these objectives can be power saving. It has been shown in recent works that structures and techniques based on cognitive radio reduce the energy consumption, while maintaining the required quality-of-service (QoS), under various channel conditions \cite{anhe1,anhe2}. Nevertheless, due to the complexity of these proposed algorithms, still vendors find it unappealing to implement these techniques. Hence, a roadway to future would be striving for more feasible, less complex, and less expensive schemes within the scope of cognitive radio.

\subsection{Cooperative Relays to deliver green communication}
\label{relay}
In infra-structured wireless networks, extending coverage of a BS is an important issue. Considering well-known properties of the wireless channel, including large path losses, shadowing effects and different types of signal fading, covering very distant users via direct transmission becomes very expensive in terms of required power in order to establish a reliable connection. This high-power transmission requirement further translates into the high power consumption and also introduces high levels of interface at nearby users and BSs.

On the other hand, in recent years, cooperative communication techniques have been proposed to create a virtual MIMO systems, where installing large antennas on small devices such as MUs is not possible. Hence, using cooperative communication, well-known improvements of MIMO systems including coverage enlarging and capacity enhancement can be achieved \cite{pabst}. Cooperative techniques also combat shadowing by covering coverage wholes \cite{pabst}. In fact, early research has shown that relaying techniques extend the battery life \cite{laneman}, which is the first step towards energy efficient networks. In particular, multi-hop communication divides a direct path between mobile terminals and BS into several shorter links \cite{Li}, in which wireless channel impairments such as path loss are less destructive, hence lower transmission power can be assigned to the BS and relays. Authors in \cite{song} mentioned that two-hop communication consumes less energy than direct communication. And finally, it has been shown in \cite{radwan} that using multi-hopping in CDMA cellular networks can reduce the average energy consumed per call.

Delivering green communication via cooperative techniques can be achieved by two different approaches. The first approach is to install fixed relays within the network coverage area in order to provide service to more users using less power. And the second approach is to exploit the users to act as relays. In this work, a relay is roughly defined as one of the network elements which can be fixed or mobile, much more sophisticated than a repeater, and it has capabilities such as storing and forwarding data, and cooperating in scheduling and routing procedures. While the second scenario eliminates the cost of installing relay nodes, it increases the complexity of the system, mostly because centralized or distributed algorithms must be designed to dynamically select relays among the users, as well as new user mobile terminals have to be designed such that they support relaying. In the two following sub-sections, we discuss these two scenarios.

\subsubsection{Enabling Green Communication via Fixed Relays}
Nonlinear signal attenuation or path loss is an interesting property of a wireless channel. This property helps to concentrate power on specific locations in a network, hence, leads to spatial reuse of various resources within a wireless network. A simple example in \cite{rost} shows that for an additive white Gaussian noise (AWGN) channel with a path loss exponent of 4, we can increase the number of BSs by a factor of 1.5 in an area unit, and reduce the transmitting power by a factor of 5, while achieving a same signal-to-noise ratio (SNR) level. In other words, a higher density of BSs leads to less energy consumption as well as a higher special reuse \cite{rost}. In fact, this is the key point which makes fixed relays a good candidate for delivering green communication as well as a general improvement of network performance. Installing new BSs in order to have a higher BS density can be very expensive. Therefore, we can install relays instead of new BSs, which is economically advantageous, and does not introduce much complexity to the network. First of all, relays need not be as high as BSs, because they are supposed to cover a smaller area with a lower power \cite{pabst}. Secondly, relays can be wirelessly connected to a BS, instead of being attached to the backhaul of the network by wire using a complicated interface \cite{pabst}. And finally, in cellular systems, unlike ad-hoc and peer-to-peer networks, complex routing algorithms are not necessary \cite{pabst}. All these reasons make installing relays a potential solution to having more energy efficient cellular networks.

In a very recent work \cite{rost}, the authors have discussed how it is possible to deliver a green communication structure in cellular networks, using fixed relays. In this paper, It is shown that relays provide a flexible way to improve the spatial reuse, are less complex than BSs and therefore cheaper to deploy, and the relays reduce the power in the system compared to systems based on direct transmission.

\subsubsection{Green Communications in Cellular Networks via User Cooperation}
User cooperation was first introduced in \cite{sendonaris}, and has been shown that not only it increases the data rate, but also the system is more robust, i.e., the achievable rates are less sensitive to channel variations. However, despite all these advantages, energy efficiency issues of user cooperation render this paradigm unappealing in wireless mobile networks. The reason is, increased rate of one user comes at the price of the energy consumed by another user acting as a relay. The limited battery life time of mobile users in a mobile network leads to selfish users who do not have incentive to cooperate. In fact, in a very recent work by Nokleby and Aazhang \cite{nokleby}, this fundamental question has been posed: whether or not user cooperation is advantageous from the perspective of energy efficiency. In this paper, a game-theoretic approach is proposed to give users incentive to act as relays when they are idle, and it is shown that user cooperation has the potential of simultaneously improving both user's bits-per-energy efficiency under different channel conditions.

User cooperation in which selfish users find cooperation favourable to their energy concerns, has recently been considered, but has still not attracted much research. However, based on existing literature, this new approach can be a promising technique to increase the system performance in terms of energy efficiency in future wireless mobile networks.

\section{Design: Addressing Energy Efficiency in Future Generation Wireless Systems}
\label{futureg}
In previous sections, we discussed that how cognitive radio and cooperative communication are becoming key technologies to address the power efficiency of a cellular network. As we mentioned earlier, European Union has already started C2POWER project with objectives to reduce power consumption of mobile terminals using cognitive and cooperative technologies by up to 50\%. In this section, we will mainly discuss techniques to enable green communication in future generation of wireless systems that will rely on cooperation and cognition to meet the increasing demand for high data rates. So far, achieving high data rate has been the primary focus of research in cooperative and cognitive radio systems, without much consideration of energy efficiency. However, many of these techniques significantly increase system complexity and energy consumption. For instance, in the context of green communication via cognitive radio, authors of {\cite{gur}} mention that there are two fundamental but entangled aspects: how to use cognitive radio for energy efficiency purposes, and how to make the cognitive radio operate in an energy efficient manner. Escalating energy costs and environmental concerns have already created an urgent need for more energy-efficient ``green" wireless communication. Hence, we need to be proactive in designing energy-efficient solutions for cooperative and cognitive networks, which will potentially drive the future generation of wireless communication. As an example, if cognitive and cooperative techniques are expected to give 50\% of power savings, then an additional 50\% improvement in the energy efficiency of these techniques will further increase the net savings by 25\%.

In the next few subsections, we will discuss an approach to obtain energy efficiency of cellular networks on an algorithmic and protocol design level, instead of energy-efficient circuitry design for communication devices.

\subsection{Low-Energy Spectrum Sensing}
The use of cognitive radio technology requires frequent sensing of the radio spectrum and processing of the sensor data which would require additional power. Therefore, it is necessary to design energy-efficient sensing schemes so that improvement in data rate due to opportunistically acquired spectrum does not lead to significant increase in the energy consumption. Low-complexity spectrum sensing techniques such as energy detection require high sensing time to accurately detect a primary signal and even fail to detect the signal at low SNR due to presence of noise-uncertainty \cite{tandra}. Therefore, detectors exploiting the cyclostationarity of the primary signals have been studied in the literature that perform better at low SNR. However, they are highly complex and need significant processing power. Therefore, design of low-complexity cyclostationary detectors needs to be investigated. Cooperative spectrum sensing improves the sensing performance by using the spatial diversity between various sensors \cite{tandra}. However, cooperative sensing would also increase the signaling overhead and thus, energy consumption. By taking into consideration the power consumed for sensing, processing and transmitting sensing data, we need to find conditions under which cooperative sensing is more energy efficient in order to achieve a certain sensing performance. New strategies should be designed to select the sensors to participate in cooperative sensing that could reduce the power consumed without severe loss in the sensing performance. Also, optimal location of the sensors should be determined that would make the sensing system energy efficient in presence of a single and multiple primary users.

Cluster-based sensing architecture has been shown to achieve higher energy efficiency and hence cluster-based designs to reduce power consumption should be considered in research \cite{wei}. Sequential detection techniques also needs to be explored to improve the energy efficiency of the system \cite{kim}. Compressive sensing has recently been proposed to reduce the complexity of wide-band sensing by sampling at a rate significantly lower than Nyquist rate, taking advantage of the sparse nature of the radio spectrum usage \cite{polo}. Therefore, efficient cooperative compressive spectrum sensing schemes is also a possible research area.

\subsection{Energy-Aware Medium Access Control and Green Routing}
Medium access control (MAC) in cooperative and cognitive wireless systems introduces a number of new challenges unseen in traditional wireless systems. For example, coordinating medium access in presence of multiple relays with different channel qualities requires a much more agile and adaptive MAC in cooperative systems. In cognitive radio systems, sensing accuracy, duration and time varying availability of primary user channels are some of the factors affecting the MAC design. The need for optimizing energy consumption further adds another dimension that can be conflicting to the goal of achieving better system performance, user satisfaction and QoS. Many of the cooperative and cognitive wireless systems will rely on multihop communication between a transmitter and its intended receiver. In addition to MAC design, proper routing schemes will thus be necessary to achieve desired end-to-end QoS.

Although, a number of MAC and routing schemes specialized for cooperative and cognitive networks exist in the literature {\cite{cormio, adam}}, little research has been done to regarding the energy efficiency of such systems. For instance, a significant volume of research exists on joint routing and spectrum allocation with objectives of throughput maximization in multi-hop cognitive and cooperative systems. In {\cite{leiding}}, decentralized and localized algorithms for joint dynamic routing, relay assignment, and spectrum allocation in a distributed and dynamic environment are proposed and analyzed. However, most of the research on joint routing and spectrum allocation does not take into account power efficiency constraints directly. Nevertheless, throughput maximization via routing-driven spectrum allocation can be interpreted as power efficiency, since more throughput is achieved using the same amount of power.

As an another example of MAC and routing schemes specialized for cooperative and cognitive networks, in \cite{alonsojournal}, Alonso-Zarate {\it et al.} proposed persistent relay carrier sensing multiple access (PRCSMA) MAC protocol employing distributed cooperative automatic retransmission request (C-ARQ) scheme (users who overheard the message can act as spontaneous relays for retransmission) in IEEE 802.11 wireless networks and in \cite{alonsoicc}, they recently evaluated the energy consumption of this protocol. In particular, Alonso-Zarate {\it et al.} described the conditions under which a C-ARQ scheme with PRCSMA outperforms non-cooperative ARQ schemes in terms of energy efficiency. On the other hand, some energy-aware MAC and routing mechanisms \cite{yahya, ergan} exist primarily for wireless sensor networks. However, sensor networks are very different than cooperative and cognitive networks in system dynamics and performance objectives. Therefore for cellular networks, objective should be to investigate novel energy-efficient MAC and routing schemes design for cooperative and cognitive wireless networks. In addition, we need to focus on optimizing energy consumption while delivering desired system performance, user satisfaction and QoS.

Hybrid-ARQ (HARQ) are another set of ARQ type protocols that use Forward-Error-Correction (FEC) coding and can be typically employed at the MAC layer to improve QoS and robustness for delay insensitive applications.  There are three important subclasses of HARQ protocols namely: HARQ-IT (Type I, in which erroneous data packets are retransmitted for memoryless detection), HARQ-CC (chase combining, where packets in error are preserved for soft combining), and HARQ-IR (Incremental Redundancy, where every retransmission contains different information bits than previous one). HARQ protocols can potentially reduce the transmission energy required for decoding at the destination for delay insensitive systems and the total energy consumption for both the transmission power and the energy consumed in the electronic circuitry of all involved terminals (source, destination and, even relays) has been studied in \cite{harq}. Hence, future MAC protocols for cognitive and cooperative systems that employ HARQ has potential to reduce energy costs of such systems.

In cooperative systems, the medium time is accessed for both direct and relayed transmissions. Addition of relayed transmission means more power consumption in the network. Future research on developing a MAC protocol that will be able to suitably quantify potential performance gain in QoS against any additional energy consumption and coordinate medium access among direct and relayed transmissions, will be important. Also, while doing so, focus should be on low-complexity schemes so that the energy savings acquired are not wasted in an increased need for processing power. For routing in a multihop cooperative system, we need to employ new protocols that can intelligently use the most energy-efficient path given the relays that are selected by the resource allocation and MAC schemes. In order to facilitate the operation of our targeted mechanisms, we must explore analytical models that can quantify trade-offs between energy savings and end-to-end QoS performance from selecting alternate routing paths. In this regard, there has already been a paradigm shift from early flooding-based and hierarchical protocols to geographic and self-organizing coordinate-based routing solutions and Internet Engineering Task Force (IETF) Routing Over Low power and Lossy networks (ROLL) working group is in process of standardization of Routing Protocol for Low power and lossy networks (RPL) \cite{rpl, ietfroll}.

For cognitive networks, energy efficiency in the MAC can be increased significantly if the access mechanism is designed to avoid collisions between primary and secondary users. Existing random access based protocols must be modified to achieve this objective in a distributed cognitive MAC with as low system complexity as possible. Statistical information of available channels can be used for QoS provisioning such as in \cite{alshamrani} but we should also consider energy efficiency as a trade-off.  Furthermore, we must focus on developing analytical models to relate important parameters of these random access methods to resulting energy consumption and QoS performance. This will enable the system engineers to choose optimal parameter values to minimize energy consumption while satisfying desired QoS performance. In a multihop cognitive radio network, due to the presence of primary user spectrum, more energy efficient routes can now be selected which would not be available without cognitive technology. Routing algorithms should be designed such that they can utilize these additional routes to minimize energy consumption.

\subsection{Energy-Efficient Resource Management with Applications in Heterogeneous Networks}
Energy consumption in wireless networks is closely related to their radio resource management schemes. Recently, power-efficient resource management for wireless networks based on cooperative and cognitive architectures has been discussed in \cite{hasan, vale}. However, current research that addresses energy efficient resource management for these systems under a variety of network objectives and constraints is not yet fully developed.

For cooperative systems, relaying mechanisms that minimize energy consumption while satisfying certain QoS performance criterion should be investigated. We also need to explore distributed schemes based on economic models with energy as a cost in the overall utility function. More specifically, we need to find answers to three fundamental questions: ``where to place relays", ``whom to relay" and ``when to relay".  In order to answer the first question ``where to place relays", we first need to obtain the optimal relay geometry, in terms of energy consumption, within a cell with different number of relays and then we must also optimize the number of relays. The second question is related to the design of optimal relay selection criterion. This relay selection criterion should be based on both fairness and energy efficiency. The authors in \cite{yweiglobecom, yweitran} have proposed an energy efficient distributed relay selection criterion with finite-state Markov channels and adaptive modulation and coding in a single user cooperative system. Such ideas should be explored for multi-user scenario. To solve the last problem of ``when to relay", design of resource allocation strategies for single and multiuser wireless systems should be studied such that relaying is selectively enabled so as to reduce overall power consumption \cite{ziaolivier, ziaicc11}.

To improve energy efficiency of cognitive radio systems, energy consumed per bit can be taken as performance metric \cite{cogenergyperbit}. We also need to investigate low power consumption based scheduling mechanisms in presence of multiple cognitive users. In \cite{hding}, the authors have proposed some energy-efficient and low complexity scheduling mechanisms for uplink cognitive cellular networks and have shown that round-robin scheduling is more energy efficient than opportunistic scheduling while providing the same average capacity and BER. Using mathematical tools based on dynamic programming and optimal control, we need to design resource allocation schemes for cognitive radio systems such that overall power consumption is minimized over a period of time while providing satisfactory performance.

Lastly, we also need to investigate into the design of energy-aware heterogeneous networks, where the macrocell (high-power node) and femtocell (low-power node) coexist for co-channel deployment. Femtocell must cognitively adapt to its surrounding environment and transmit in such a way in order not to create cross-tier interference to main cellular network \cite{bharucha}. Using cooperative relaying, the network coverage of these femtocells can be improved without causing huge interference to the macrocell system. For such heterogeneous network architecture, employing power-efficient resource management techniques such as selective relaying, energy efficient modulation etc. can be very attractive. Further research on optimal cell sizes and femtocell BS locations taking into consideration the energy spent for the system backhaul and signaling overhead, can save a lot of power. This way we can further reduce the energy consumption of macrocell BSs and user handsets to achieve a given system performance.

\subsection{Cross-Layer Design and Optimization}
Cellular wireless communication systems ought to support different kinds of applications, including voice and data applications. Each application has a different energy limit, required bit rate, bit error rate, delay constraints, outage probability, etc. Traditionally, these requirements have been tried to be achieved within the scope of a layered structure, called protocol stack. In fact, there has been a significant amount of research tackling these problems within each layer, assuming each layer operates independently of the other layers. Examples of these efforts in the link layer are employing MIMO techniques, channel coding, power control and adaptive resource allocation techniques. In the MAC layer, different channelization or random access schemes, along with scheduling and power control can be mentioned. Moreover, regarding the network layer, a rich literature can be found on the energy-constrained and delay-constrained routing. And finally, adjusting QoS requirements adaptively is a venue to meet the users demand in the application layer \cite{goldsmith}.

The rationale behind using the protocol stack is that it helps the designer to break the design problem into several simpler problems, namely layer modules. This paradigm also makes the evaluation of the proposed algorithms easier. However, limiting each layer to be independent of others and sub-optimality of this modularized paradigm lead to a poor performance, especially when the resources such as energy are scarce \cite{goldsmith}. Therefore, cross-layer design can be a very useful tool to minimize energy consumed across the entire protocol stack. In cross-layer design, we try to escape from the limitations that the traditional waterfall-like concept of protocol stack imposes. In the new paradigm, we want to not only consider the interdependencies between different layers, but also take advantage of them. In particular, for energy saving purposes, it is necessary to consider the invariably changing operation conditions in cellular networks. Due to the mobility of the users, and also characteristics of wireless channel along with the nature of modern applications, propagation environment and application requirements are time varying. Thus, more holistic control algorithms from cross-layer perspective must be designed which adapts the system to these dynamics at run time. Going in this direction, greener cellular communication systems can be delivered compared to the existing ones \cite{dejonghe}. However, if not carefully designed cross-layer design might itself lead to increased complexity and energy consumption. Hence, we should explore the cross-layer alternatives to schemes proposed for an individual layer (PHY, MAC etc.) and analyze important tradeoffs in energy consumption and system performance. Multiple relays in cooperative communication and spectrum sensing mechanism in cognitive radio networks introduce new challenges in cross-layer design for these networks \cite{vousoughi, luo}.

One of the objectives should be to devise cross-layer schemes that will allow joint optimization of some or all of the following parameters: assigning the subcarriers, rates, and power (physical layer attributes), channel access mechanisms (MAC layer attribute), routing (network layer attribute), and rate (transport layer attribute) while taking into account system related errors (e.g. sensing errors in cognitive radios) and other errors that contribute to cross layer issues. Further, in order to save energy in an ad-hoc wireless network, more packets should be transmitted when channel quality is good and at the same time collision losses due to network congestion must be reduced so that the packets need not to be retransmitted. In this regard, we need to explore new mechanisms for cognitive radio and cooperative ad-hoc networks by which channel traffic can be measured either by single-bit or multi-bit signalling overhead or by loss or delay in the network. Energy-efficient cross-layer schemes that will optimize resources in various layers while considering the channel quality as well as the network traffic, should be investigated.

An example, where cross-layer design would be crucial, is in the use of cooperative relaying to improve spectrum diversity in cognitive radio networks \cite{qzhang}. Large gains in efficiency and fairness of resource sharing can be obtained by cooperation among cognitive radio nodes. Specifically, some cognitive radio users with low traffic demand can help improve spectrum efficiency by acting as relays for the cognitive radio users that have high traffic demand but low available bandwidth. For such a network, cross-layer design is important as while performing resource allocation (relay, power etc.), transmission demand of each cognitive radio user has to be taken into account.

\subsection{Addressing Uncertainty Issues}
Most research in the field of cooperative and cognitive radio systems is mainly based on the assumption of perfect channel state information (CSI), which is often unrealistic in practice. Presence of non-Gaussian noise, quantization effects, fast varying environment, delay in CSI feedback systems and hardware limitations are the main factors that cause errors in CSI. The performance of cognitive radio sensing system is drastically impaired, when various wireless channels e.g., detecting, reporting, and inter-user channels have uncertainty \cite{ywang}. For cooperative systems, the optimal relay selection and robust resource allocation with imperfect CSI has also remained largely unexplored. For cognitive radio systems, it is also important to take into account the effects of imperfect sensing \cite{ruan}. Spectrum sensing is further complicated due to uncertainty in interference from other secondary networks \cite{ghasemi}. Providing robustness in conjunction with energy efficient solutions to such scenarios is, therefore, a task of significant practical interest. Also, the robustness of the efficient scheduling schemes for MAC and cross-layer optimization needs further investigation taking uncertainty in the channel congestion into account. In order to maintain energy savings under imprecise conditions, we must investigate the robustness of our proposed energy-efficient schemes and compare the performance with the existing schemes in practical scenario with uncertain environment. Hence, depending on the QoS targets, robust algorithms for energy efficient resource optimization considering uncertainty in CSI, should be explored.

\section{Some Broader Perspectives}
\label{broader}
The most important issue in developing networks which are energy-aware is to model the consumption of the wireless interfaces \cite{anastasi}. Usually, the wireless interface consumes energy with the same rate in receive, transmit or idle states. In turn, the less the wireless interface is operating, the less energy is consumed. Based on the preceding argument, the best strategy to minimize the energy consumption is to shut down the wireless interface, or to go to energy saving mode as much as possible. In order to achieve this, algorithms needed to determine when it is suitable to switch to energy saving mode or turning off the transceivers. We already have discussed strategies with the aforementioned concept in this paper. For instance, we have mentioned Discontinuous Reception (DRX) and Discontinuous Transmission (DTX) modes in LTE standard, and sleep mode mechanism in IEEE 802.16e, both for mobile terminals. We also have talked about enabling sleep mode for BSs. However, these methods are based on instantaneous observations. On the other hand, the traffic pattern is dramatically different in different times of the day or in different geographical locations. In a broader perspective, there can be a data-base in BS and mobile terminals, in which the traffic pattern during different times of the day is saved. Based on this obtained statistics, dynamic algorithms can be designed in order to switch the BS or mobile terminal to a different power profile appropriate for that time of the day. In a recent paper by Dufkova {\emph{et al.} \cite{dufkova}}, it has been shown that if such predictions on users are available, savings from $25\%$ up to $50\%$ can be achieved, depending on the time of the day. However, their results are based on off-line optimizations and represent an upper bound on the energy savings possible.

From another perspective, BSs distributed over a certain geographical area are connected to a power grid. In recent years, smart grid has emerged to coordinate the power generators, transmission systems and appliances utilizing two-way communication lines between all these different entities. These two-way communication lines can be dedicated point-to-point wireless channels or IP-based connections \cite{ykim}. On the other hand, BSs in general, are power hungry elements. Hence, looking at BSs as power consumers or appliances, and absorbing them in a smart grid can exceedingly increase the power efficiency without adversely affecting the QoS and capacity. This can be done by adding measurement sensors which can update the status of BSs, and then transmit them to the other BSs and smart grid control system. Here, in addition to cooperation with each other, BSs also cooperate with the power system to manage the energy consumption.

Contrary to the ideas mentioned by far, Humar {\emph{et al.} in \cite{humar}} suggest a different way of thinking in energy efficiency modeling. Almost all the research on making cellular communication green results in larger number of BSs with lower level of powers, since the objective is to reduce the {\emph{operating energy}}. However, authors of {\cite{humar}} noticed that in all the cases, the new BSs are more sophisticated equipments, and producing these sophisticated equipments requires more energy compared to conventional ones. This energy which is associated with all the processes of producing an equipment is called {\emph{embodied energy}}. According to this paper, embodied energy accounts for a significant proportion of energy consumed by the BS, and taking this energy into account along with operating energy in modeling cellular network's energy consumption results in solutions which disagree with increasing the number of BSs and lowering their power.

\begin{center}
\begin{table}[ht]
\caption{Energy savings obtained by some of the discussed techniques}
{\small
\hfill{}
       \begin{tabular}{ p{4cm}  p{4cm}}
    \hline
    \bfseries Description     &     \bfseries Reported savings \\ \hline
    Improvements in Power Amplifier & - up to 50\% with doherty architecture and GaN-based amplifiers \\
                                    		    &  - up to 70\% with switch-mode power amplifiers  \\  \hline
    Network self-organizing techniques	& between 20-40\% BS power savings \\ \hline

    Renewable Energy Resources in off-grid sites &   up to 0.35\% of global diesel consumption	\\ \hline
    Heterogeneous network deployment &   up to 60\% savings compared to a network with macro-cells	\\ \hline
    Dynamic spectrum management & up to 50\% 	    \\ \hline

    \end{tabular}
}
\hfill{}
\label{tb:energysavings}
\end{table}
\end{center}

\section{Conclusion}
\label{conc}
This paper addresses the energy efficiency of cellular communication systems, which is becoming a major concern for network operators to not only reduce the operational costs, but also to reduce their environmental effects. We began our discussion with green metrics or energy efficiency metrics. Here, we presented a brief survey of current efforts for the standardization of the metrics and the challenges that lay ahead. Regarding architecture, since BSs represent a major chunk of energy consumed in a cellular network, we then presented an exhaustive survey of methods that have been currently adopted or will be adopted  in future in order to obtain energy savings from BSs. In particular, we discussed the recent improvements in power amplifier technology that can be used to bring energy savings in BSs. Improvements in the power amplifier will not only decrease the power consumption of the hardware system, but will also make the BS less dependant on air-conditioning. We also discussed the power saving protocols such as sleep modes, that have been suggested for next generation wireless standards. Such power saving protocols at the BS side still need to be explored in future wireless systems. Next, we discussed energy-aware cooperative BS power management, where certain BSs can be turned off depending on the load. A recent concept called ``Cell zooming" appears to be a promising solution in this regard. Another way to significantly reduce the power consumption of BSs, in particular, those at the off-grid sites, is by using renewable energy resources such as solar and wind energy in place of diesel generators. Lastly, we discussed how minimizing the number of BSs with a better network design and bringing minor architectural changes can be beneficial in achieving energy efficiency.

Heterogeneous network deployment based on smaller cells such as micro, pico and femtocells is another significant technique that can possibly reduce the power consumption of a cellular network. However, as some of the recent research suggests, careful network design is required as deploying too many smaller cells may in fact reduce the power efficiency of the central BS. Also, when a large number of BSs with small cell sizes are deployed, the embodied energy consumption will dominate and lead to an increase in total energy consumption \cite{humar}.  We also discussed how emerging technologies such as cognitive radio and cooperative relaying can be useful for obtaining ``green" network technology. In this regard, we discussed research challenges to address energy efficiency in cognitive and cooperative networks including low-energy spectrum sensing, energy-aware MAC and routing, efficient resource management, cross-layer optimization, and uncertainty issues. Finally, we explored some broader perspectives such as statistical power profiles, smart grid technology and embodied energy to achieve energy efficient cellular network. Table \ref{tb:energysavings} lists the energy savings reported by authors, that can be obtained by some of the techniques discussed in the paper.

In summary, research on energy efficient or ``green" cellular network is quite broad and a number of research issues and challenges lay ahead. Nevertheless, it is in favor of both the network operators and the society to swiftly address these challenges to minimize the environmental and financial impact of such a fast growing and widely adopted technology. This article attempts to briefly explore the current technology with respect to some aspects related to green communications and we discuss future research that may prove beneficial in pursuing this vision.

\section*{Acknowledgment}
This research was supported by the Natural Sciences and Engineering Research Council of Canada (NSERC) under their strategic project award program and in part by Alexander Graham Bell Canadian Graduate Scholarship.

\end{document}